\documentclass[]{article}

\newcommand{\ga}{\alpha} \newcommand{\gb}{\beta}
\newcommand{\gc}{\gamma} \newcommand{\gd}{\delta}

\newcommand{\sbs}{\subseteq} 
 \newcommand{\lto}{\rightarrow}

\newcommand{\sfc}{{\sf C}}\newcommand{\sn}{{\sf N}}

\newcommand{\ca}{{\cal A}} \newcommand{\cl}{{\cal L}}
\newcommand{\cb}{{\cal B}}

\newcommand{\vd}{\vdash}

\newcommand{\Cn}{{\hbox{\rm Cn}}}
\newcommand{\Coh}{{\hbox{\rm Coh}}}

\newtheorem{theorem}{\bf Theorem}
\newtheorem{lemma}[theorem]{\bf Lemma}

\newtheorem{corollary}[theorem]{\bf Corollary}

\newenvironment{definition}{\par\medskip\addtocounter{theorem}{1}%
  \noindent{\bf Definition \arabic{theorem}}\ }{\medskip}

\newcommand{\qed}{\vrule height5pt width3pt depth0pt}

\newcommand{\ev}{\makebox[1.3em]{$\rule{0.1mm}{2.5mm}\hspace{-1.2mm}\sim$}}
\newcommand{\notev}{\makebox[1.3em]{$\rule{0.1mm}{2.5mm}\hspace{-1.2mm}\not\sim$}}

\newcommand{\prev}{\preceq}
\newcommand{\notprev}{\not\preceq}
\newcommand{\sev}{\makebox[1.8em]{$\rule{0.1mm}{2.5mm}\hspace{-1.3mm}\sim_\preceq$}}

\newcommand{\wev}{\makebox[1.8em]{$\rule{0.1mm}{2.5mm}\hspace{-1.3mm}\sim^w_\preceq$}}

\title{Entrenchment Relations: A Uniform Approach to Nonmonotonicity\thanks{Work
supported by Training through Research Contract No. ERBFMBICT950324 between
the European Community and Universit\`{a} degli Studi di Roma ``La
Sapienza''.}}
\author{Konstantinos Georgatos\\ Dipartimento di Informatica e Sistemistica\\
Universit\`a di Roma ``La Sapienza''\\ Via Salaria 113, Roma 00198\\ Italy}

\date{}
\begin{document}
\maketitle

\begin{abstract}
We show that Gabbay's nonmonotonic consequence relations can be reduced to a new
family of  relations, called entrenchment relations. Entrenchment
 relations provide a direct generalization of  epistemic
entrenchment and expectation ordering introduced by G\"ardenfors and Makinson for the
study of belief revision and expectation inference, respectively.
\end{abstract}

\section{Introduction}

Nonmonotonicity has offered great promise as a logical foundation for knowledge
representation formalisms. The reason for such a promise is that nonmonotonic logic
allows ``jumping'' to conclusions, completes in a reasonable way our (incomplete)
knowledge and withdraws conclusions in the light of new information. Therefore, most
approaches to central problems of Artificial Intelligence, such as belief revision,
database updating, abduction and action planning, seem to rely on one way or another to
some form of nonmonotonic reasoning.

There are several proposals of logical systems performing nonmonotonic inference.
Among  the most popular of them are: circumscription, negation as failure, default
logic, (fixed points of) various modal logics and inheritance systems.  However, and
despite the numerous results intertranslating one of the above systems to the other,
none of the above formalisms emerged as a dominant logical framework under which all
other nonmonotonic formalisms can be classified, compared and reveal their logical
content. This fact  signifies that our intuitions on the
process of nonmonotonic inference are fragmented. Although, all the above mentioned
logics are worth be studied and employed as a central inference mechanism they cannot
serve as the place where finally our basic intuitions about nonmonotonicity can rest.

Addressing this problem, Gabbay in \cite{GAB85} proposed to study nonmonotonic
inference through Gentzen-like context sensitive sequents. Following this proposal, a
new line of research flourished by studying properties of the so-called {\em
nonmonotonic consequence relations} leading to a semantic characterization through (a
generalization of) Shoham's preferential models. This line of research led to
classification of several nonmonotonic formalisms and recognized several
logical properties properties that a nonmonotonic system should desirably satisfy
such as cumulativity or distributivity. However, there are two disadvantages of this
framework:
\begin{itemize}
\item nonmonotonic consequence relations express the sceptical inference of a
nonmonotonic proof system and therefore fail to describe nonmonotonicity in its full
generality, that is, the existence of multiple extensions.
\item it does not seem that there is a straightforward way to design a nonmonotonic
consequence relation from existing data unless they already encode some short of
conditional information (see~\cite{LM92}).
\end{itemize}
 These two disadvantages suggest that a nonmonotonic consequence relation is not a
primitive notion but {\em derived} from a more basic inference mechanism.

In this paper, we shall introduce a novel framework for generating nonmonotonic
inference, through a class of   relations, called {\em
entrenchment  relations\/}. We shall see that the framework of entrenchment
 relations is at least as expressive than that of nonmonotonic consequence
relations. In particular, nonmonotonic consequence relations can be {\em reduced} to
entrenchment  relations (in the classical case) while the inference defined
through entrenchment  relations admits and identifies the existence of multiple
extensions. On the other hand, entrenchment relations seem to build inference
easily and from the bottom up. Simple frequency data, from example generates easily at
least one class of them (rational orderings --- see~\cite{AG96}).

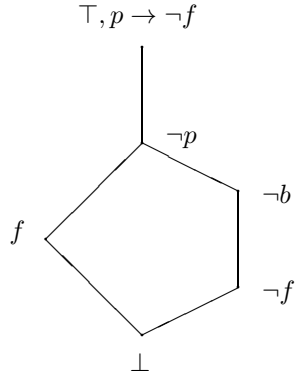
\begin{figure}[h]
\centering
\setlength{\unitlength}{0.00083300in}%
\begin{picture}(1575,2346)(1576,-2113)
\thinlines
\put(2401,-61){\circle*{.3}}
\put(2401,-61){\line( 0,-1){600}}
\put(2401,-661){\line( 2,-1){600}}
\put(2401,-661){\circle*{.3}}
\put(3001,-961){\line( 0,-1){600}}
\put(3001,-961){\circle*{.3}}
\put(3001,-1561){\line(-2,-1){600}}
\put(3001,-1561){\circle*{.3}}
\put(2401,-1861){\line(-1, 1){600}}
\put(2401,-1861){\circle*{.3}}
\put(1801,-1261){\line( 1, 1){600}}
\put(1801,-1261){\circle*{.3}}
\put(2000, 89){$\top, p \lto\neg f$}
\put(2551,-661){$\neg p$}
\put(3151,-1036){$\neg b$}
\put(3151,-1636){$\neg f$}
\put(2326,-2086){$\bot$}
\put(1576,-1261){$f$}
\end{picture}
\caption[]{A (transitive) entrenchment relation.}{\label{fig:tree}}
\end{figure}

Entrenchment relations are  relations and will be denoted by $\prev$. $\ga\prev\gb$
will be read as
\begin{quote}
$\gb$ is at least as entrenched as $\ga$
\end{quote}
in the sense that ``$\ga$ is more defeasible than $\gb$''. In other words, ``if $\ga$
is accepted then so is $\gb$''. For example, consider the  partial description of a
(transitive) entrenchment
 relation in Figure~\ref{fig:tree}.

In that figure,
a path upwards from $\ga$ to $\gb$ indicates that $\ga\prev\gb$, where $\prev$
denotes the entrenchment relation.
The entrenchment relation of Figure~\ref{fig:tree} says for example that
$\bot$ is less entrenched than all formulas, $\neg f$ is less entrenched than
$b$,
$p$ and
$p\land f$, and $f$ is less entrenched than  $p$.

How will an entrenchment relation be used for inference? The idea is simple. We shall
use  entrenchment for {\em excluding} sentences.
\begin{quote}
 A sentence $\ga$ will infer
(in a nonmonotonic way) another sentence $\gb$ if $\ga$ together with a sentence
$\gc$, {\em that is not less entrenched than} $\neg\ga$, (classically) imply $\gb$.
\end{quote}
The reason we exclude sentences less entrenched than $\neg\ga$ is that if we allow
such a sentence then we should also allow $\neg\ga$. However this will bring
inconcistency.
 For instance, using the above example  and
assuming
$p$ we should exclude $\neg p$,
$\neg b$  and
$\neg f$. We remain with
$p\lto\neg f$. Adding $p\lto\neg f$ to the classical theory of $p$ we have that $p$
nonmonotonically implies $\neg f$. Similarly, assuming $b$ we exclude $\neg b$ and
$\neg f$. We remain with $\neg p$ and $f$. So, $b$  nonmonotonically implies
$f$ and $\neg p$. With no assumptions we have two consistent sets of sentences that
remain after excluding $\bot$: $\{p\lto \neg f, \neg p, f\}$ and $\{p\lto \neg f,
\neg p, \neg b, \neg f \}$. Therefore, it is possible to have more than one
possibilities for extending the theory of our assumptions and that leads to the
well-known phenomenon of multiple extensions.

It is clear that our framework separates nonmonotonic inference to two different
monotonic proof procedures: one positive and the other negative. The claim that
entrenchment relations is a useful concept towards our understanding of nonmonotonicity
will be substantiated by a series of representation results. We shall show that
Gabbay's nonmonotonic consequences relations can be expressed through entrenchment
relations, and identify those classes of entrenchment relations which correspond to the
classes of nonmonotonic consequence relations that have attracted special interest in
the literature. In addition our
framework provides more:

\begin{itemize}
\item {\em Uniformity}. Inference defined through an entrenchment relation remains the
same throughout the above characterization.
\item {\em Monotonicity\/}. Any strong cumulative nonmonotonic inference relies on a
monotonic (on both sides) entrenchment relation.
\item {\em Identification of multiple extensions\/}. The way we define inference allows the
identification of multiple extensions. Therefore, both the sceptical and  credulous
approach towards nonmonotonicity are expressible in our framework.
\item {\em A conceptually primitive view of nonmonotonicity\/}. In our framework, a
nonmonotonic formalism separates into two logical mechanisms handling positive and
negative information.
\end{itemize}

Entrenchment relations provide a generalization of G\"ardenfors-Makinson's
expectation orderings introduced for the characterization of expectation nonmonotonic
consequence relations (\cite{GM94}). This result was later extended to rational
consequence relations in~\cite{KG96e}. In~\cite{FHL94}, incompletely specified
expectation orderings were studied. But, to our knowledge, there is no study of such
relations outside the non-Horn classes of nonmonotonic consequence relations. This
paper fills exactly this gap by showing that {\em all} nonmonotonic consequence
relations can be represented through entrenchment relations.

However, entrenchment relations have a close relative in the study of belief
revision, called epistemic entrenchments~(\cite{GM88}). Epistemic
entrenchments proved to be a very useful for belief revision and became the
standard tool~(\cite{}) for studying the AGM postulates~(\cite{AGM85}).
Moreover, generalizations of epistemic entrenchment have been proposed by
Lindstr\"{o}m-Rabinowicz~(\cite{LR91}) and Rott~(\cite{ROT92}). They both
proposed to drop linearity from epistemic entrenchment.
Lindstr\"{o}m-Rabinowicz used such a partial ordering for the study of
relational belief revision. On the other hand, Rott's {\em generalized
epistemic entrenchments}, use the original G\"ardenfors-Makinson syntactic
translation for generating belief revision functions. As a consquence, Rott
characterizes non-Horn belief revision functions with Horn epistemic
entrenchments and vice versa. Therefore, our results cannot be derived, even
through a suitable translation, by the above works, although intuition and
motivation should be credited on both of them.

Relations of expectation orderings with other systems performing some sort of
nonmonotonic reasoning are abundant
(\cite{BOU92a},\cite{BOU92b},\cite{LAM91},\cite{LAM92},\cite{WOB92},\cite{DUP91}),
such as Pearl's system $Z$ (\cite{PEA90}), conditional logic
(\cite{STA68},\cite{LEW73}), and possibilistic logic (\cite{DUP88}). It is worth
mentioning that orderings  appear abundantly in the
literature of nonmonotonic logic. Orderings of models lead to the preferential model
framework (\cite{SHO88},\cite{KLM90},\cite{KS91},\cite{MAK94}), while ordering of
sentences lead to {\em prioritization}. Most nonmonotonic formalisms have been
enriched with priority handling. However, entrenchment relations are not priorities
but rather {\em rules} for extending a special form of priority statements. The
connection of priority statements with entrenchment relations are similar to that of
sequents with proof rules.

The further contents of this paper are as follows. In Section 2, we present
entrenchment relations. We discuss their informal meaning  and present
various properties of them. Then, in Section 3, we define the notion of
maxiconsistent and weak maxiconsistent inference as derived from a pair of
relations. Both inference schemes can generate all nonmonotonic consequence
relations. In Section 4, we review nonmonotonic consequence relations and
present our representation results. In Section~\ref{section:summary}, we
summarize.

\section{entrenchment Relations}
\label{sec:ccr}

G\"ardenfors and Makinson recently showed  (\cite{GM94}) that the study of a strong
non-Horn class of nonmonotonic consequence relations, called {\em entrenchment
inference relations\/}, can be reduced to the study of a particular class of linear
preorders among sentences called {\em entrenchment orderings}.
Subsequently, in~\cite{KG96e}, the author  extended this result to the
well-known Lehmann and Magidor's class of rational inference relations (\cite{LM92}).
The purpose of this paper is to show that the study of {\em all\/}  nonmonotonic
consequence relations can be reduced to the study of relations among sentences which
generalize the class of above mentioned orderings.

The interpretation  of entrenchment
orderings which G\"ardenfors and Makinson proved equivalent to entrenchment inference
relations is the following. Assume there is an ordering $\prev$ of the sentences of a
propositional language $\cl$, where $\ga\prev\gb$ means ``$\gb$ is at least as entrenched
as $\ga$'' or ``$\gb$ is at least as surprising as $\ga$''.  Therefore, $\prev$ is a
relation comparing degrees of defeasibility among sentences.

This interpretation of $\prev$, as well as a similar one based on
possibility given in~\cite{FHL94}, although seems fit
for the particular class of nonmonotonic inferences it characterizes, has in our
opinion the following disadvantages. First, it has a complicated flavor by relying on
notions such as expectation, defeasibility, and surprise that are far from primitive.
Second, it points to a semantical interpretation by committing to a subjective
evaluation of sentences and therefore is lacking the proof-theoretic interpretation
meant for relations generating inference. Finally, this interpretation loses its
plausibility once we weaken one of its defining properties (for example linearity or
transitivity) and restricts us to a unique class of orderings.

Entrenchment relations are nothing more than a generalization of the above ordering.
We will drop first linearity of the preorder, for characterizing preferential
inference, and subsequently transitivity.
Note that the entrenchment  interpretation  is weakened  once we drop transitivity: if
a sentence
$\ga$ is less entrenched than
$\gb$ and $\gb$ less entrenched than $\gc$, then $\ga$ should be less entrenched than
$\gc$. However, inference through an entrenchment relation remains the same, that is we
still exclude sentences that relate to $\neg\ga$, i.e. $\gb\prev\neg\ga$. Therefore
the notion of entrenchment becomes contextual.  The situation is similar to that of a
consequence relation that it is not necessarily monotonic, that is, just as
$\ga\ev\gc$ but not neseccarily $\gb\ev\gc$ whenever $\gb\vd\ga$. However, our
representation remains useful as we can express multiple extensions.

Here are our assumptions on the language. We  assume a language
$\cl$ of propositional constants closed under the boolean connectives $\lor$
(disjunction),
$\land$ (conjunction),
$\neg$ (negation) and $\to$ (implication). We use greek letters  $\ga$, $\gb$, $\gc$,
etc. for propositional variables. We also assume an underlying consequence relation
that it will act as the underlying proof-theoretic mechanism.  For all practical
purposes, it can be thought as classical propositional calculus, but all following
definitions and theorems can be carried out in  any consequence relation
$\vd\sbs 2^{\cl}\times\cl$ that includes  classical propositional logic, satisfies
compactness (i.e., if
$X\vd\gb$ then there exists a finite subset $Y$ of $X$ such that
$Y\vd\gb$)\footnote{We write
$X,\ga\vd\gb$ for $X\cup\{\ga\}\vd\gb$.}, the deduction theorem (i.e., $X,\ga\vd\gb$
if and only if $X\vd\ga\to\gb$) and disjunction in premises (i.e., if $X,\ga\vd\gb$
and
$X,\gc\vd\gb$ then $X,\ga\lor\gc\vd\gb$). We denote the consequences of $\ga$ with
$\Cn(\ga)$. We should add that nonmonotonic inference which does not contain
classical tautologies is a rather rare exemption.

Now, let us assume a relation
$\prev$ between sentences of $\cl$. $\ga\prev\gb$ should be interpreted as
\begin{quote}
$\gb$ is (at least) as entrenched as  $\ga$.
\end{quote}
Now, read $\prev$ as depending on $\gb$, that is as a unary predicate
indexed by $\gb$. Therefore,  if we strengthen the left part we
expect this relation to hold. On the other hand, sentences on the right of $\prev$
express {\em context\/}, so properties imposed on that part translate to our
conception of context. We can be either monotonic or non-monotonic on context. We will
see that either way can still generate nonmonotonic inference. What then would the
properties of
$\prev$ be? We shall assume the following three basic properties:
\begin{enumerate}
\item $\ga\prev\ga$ \qquad (Reflexivity)
\item If $\ga\vd\gb$ and $\gb\prev\gc$ then $\ga\prev\gc$. \qquad (Left Monotonicity)
\item If $\ga\vd\gb$ and $\gb\vd\ga$ then $\gc\prev\ga$ iff $\gc\prev\gb$. (Logical
Equivalence)
\end{enumerate}

The meaning of Reflexivity is straightforward.

Left Monotonicity  says that if $\gb$ is less entrenched then $\gc$ so is any sentence
stronger than
$\gb$.

Finally, Logical Equivalence says that two logically equivalent sentences (under
$\vd$) construct the same context and therefore if a sentence is less entrenched than
one of them must be less entrenched than the other as well.

 We summarize the above in the
following definition of entrenchment frame.

\begin{definition} An {\em entrenchment frame\/} is a pair $\langle \cl, \vd, \prev
\rangle$, where
$\prev$ is a  relation on $\cl\times\cl$, called  {\em entrenchment
 relation\/}, that satisfies the above properties, that is Reflexivity,
Left Monotonicity and Logical Equivalence.
\end{definition}

All properties of entrenchment relations mentioned in the subsequent appear on
Table~\ref{table:properties}.
\begin{table}[p]
\begin{center}
\begin{minipage}{4.5in}
{\small
\begin{tabular}{cl}

$\ga\prev\ga$ & {\footnotesize (Reflexivity)} \vspace{.1in} \\

$\displaystyle \frac{\ga\vd\gb \qquad \gb\prev \gc }{\ga \prev \gc}$  &
{\footnotesize (Left Monotonicity)} \vspace{.1in}\\

$\displaystyle \frac{ \ga\vd\gb }{\ga \prev \gb}$ & {\footnotesize(Dominance)}
\vspace{.1in}\\

$\displaystyle \frac{ \ga\vd\gb \qquad \gb\vd \ga \qquad \gc\prev\ga}{\gc\prev\gb}$ &
{\footnotesize(Logical Equivalence)} \vspace{.1in}\\

$\displaystyle \frac{\ga\lor\gb\prev\gb \qquad \ga\lor\gb\prev \ga \qquad
\ga\lor\gc\prev\ga}{\gb\lor\gc\prev\gb}$ & {\footnotesize(Weak Equivalence)}
\vspace{.1in}\\

$\displaystyle \frac{\ga\prev\gb \qquad \gb\prev \ga \qquad
\gc\prev\ga}{\gc\prev\gb}$ & {\footnotesize(Equivalence)} \vspace{.1in}\\

$\displaystyle  \frac{\ga\lor\gb\prev\ga \qquad
\ga\lor\gc\prev\ga}{\ga\lor\gb\lor\gc\prev\ga}$   & {\footnotesize(Weak Left
Disjunction)} \vspace{.1in}\\

$\displaystyle  \frac{\gb\prev\ga \qquad
\gc\prev\ga}{\gb\lor\gc\prev\ga}$  & {\footnotesize(Left
Disjunction)}\vspace{.1in}\\

$\displaystyle  \frac{\ga\lor\gb\lor\gc\prev\ga\lor\gb \qquad
\ga\lor\gb\prev\ga}{\ga\lor\gc\prev\ga}$  & {\footnotesize(Weak Bounded
Cut)}\vspace{.1in}\\

$\displaystyle  \frac{\gc\prev\ga\lor\gb \qquad
\gb\prev\ga}{\gc\prev\ga}$  & {\footnotesize(Bounded  Cut)}\vspace{.1in}\\

$\displaystyle  \frac{\ga\lor\gc\prev\ga \qquad
\ga\lor\gb\prev\ga}{\ga\lor\gb\lor\gc\prev\ga\lor\gb}$ & {\footnotesize(Weak Bounded
Right Monotonicity)}
\vspace{.1in}\\

$\displaystyle  \frac{\gc\prev\ga \qquad
\gb\prev\ga}{\gc\prev\ga\lor\gb}$ & {\footnotesize(Bounded Right Monotonicity)}
\vspace{.1in}\\

$\displaystyle  \frac{\ga_0\prev\ga_n \quad\ga_n\prev\ga_{n-1}\quad \cdots
\quad\ga_1\prev\ga_0 }{\ga_n\prev\ga_0}$ & {\footnotesize(Acyclicity)}
\vspace{.1in}\\

$\displaystyle  \frac{\ga_0\lor\ga_1\prev\ga_0 \quad\ga_1\lor\ga_2\prev\ga_1\quad
\cdots\quad
\ga_n\lor\ga_0\prev\ga_n }{\ga_0\lor\ga_n\prev\ga_0}$ &
{\footnotesize(Weak Acyclicity)}
\vspace{.1in}\\

$\displaystyle  \frac{\ga\prev\gb \qquad
\gb \vd \gc}{\ga \prev \gc}$ & {\footnotesize(Right Monotonicity)} \vspace{.1in}\\

$\displaystyle  \frac{\gc\prev\ga \qquad
\gc\prev\gb}{\gc\prev\ga\land\gb}$   & {\footnotesize(Right Conjunction)}
\vspace{.1in}\\

$\displaystyle  \frac{\ga\prev\gb \qquad
\gb\prev\gc}{\ga\prev\gc}$ & {\footnotesize (Transitivity)} \vspace{.1in}\\

$\ga\prev\gb$ or $\gb\prev\ga$ & {\footnotesize(Connectivity)}

\end{tabular}
}

\end{minipage}

\end{center}

\caption{\label{table:properties} Properties for entrenchment relations}
\end{table}

The following property has been considered in the framework of entrenchment  orderings
(\cite{GM94})
\begin{quote} If $\ga\vd\gb$ then $\ga \prev \gb$.\quad (Dominance)
\end{quote} In view of Reflexivity,  Left Monotonicity implies Dominance. Given
Dominance and Reflexivity of $\vd$, Reflexivity of $\prev$ follows. Dominance is a
very useful property that is used abundantly in the subsequent and was in fact one of
the defining properties of G\"ardenfors and Makinson's entrenchment ordering and
epistemic entrenchment.

The following property is derived by Left Monotonicity
\begin{quote}
$\ga\vd\gb$ and $\gb\vd\ga$ implies $\ga\prev\gc $ iff
$\gb\prev\gc$.
\end{quote}

While Left Monotonicity allows us to strengthen arbitrarily
sentences on the left, Bounded Left Disjunction and
Left Disjunction  allow us to weaken them. These properties amount to a disjunction
property. An entrenchment relation
will be called  {\em disjunctive (weak disjunctive)\/} if it satisfies Left
Disjunction (Weak Left Disjunction). Similarly, for entrenchment frames.

The definition of entrenchment frame says nothing about how one should go combining sentences
on the right, i.e., combining contexts.
Bounded  Cut and Right
Conjunction express our ability to  strengthen  the right part so strengthen  the
context. Bounded Right Monotonicity, Right Monotonicity and Weak Right Monotonicity
weaken the right part so weaken the context. It is worth noting that Right
Conjunction makes a sentence, less entrenched than another sentence and its negation,
less entrenched than all sentences. Right conjunction allow us to combine
contexts using conjunction. Bounded  Right Monotonicity follows from Right
Monotonicity. Weak Bounded Right Monotonicity together with Bounded Cut implies Weak
Left Disjunction.

Weak Bounded Right Monotonicity and Weak Bounded Cut together  are equivalent to Weak
Equivalence.

Bounded Cut and Bounded Right Monotonicity together imply Equivalence. While given
Left Disjunction, Equivalence implies Bounded Cut and Bounded Right Monotonicity.

Observe that Transitivity implies Right Monotonicity, and thus Bounded Right
Monotonicity. Transitivity is equivalent to Right Monotonicity given Bounded Cut.
Transitivity and Dominance implies Left Monotonicity.

Connectivity is the only non-Horn property among the above properties.
Therefore, any class of entrenchment connectivity relations satisfying the above
properties except Connectivity is closed under intersections. G\"ardenfors and
Makinson merged Connectivity and Right Conjunction into

\begin{quote}
 $\ga\prev\ga\land\gb$ or $\gb\prev\ga\land\gb$.\quad (Conjunctiveness)
\end{quote}
and along with Dominance and Transitivity make the defining set of properties of
G\"ardenfors and
Makinson's entrenchment orderings which is the notion we generalize.

In this paper, we will only
study Horn properties. To our knowledge previous results concern only non-Horn entrenchment
 relations satisfying connectivity: entrenchment and rational ordering in
\cite{GM94} and
\cite{KG96e}, respectively.

\section{Maxiconsistent and Weak Maxiconsistent Inference}
\label{sec:max-inference}

We shall now describe an inference scheme based on an entrenchment relation $\prev$. We
will  define two finitary consequence relation, that is  subsets of
$\cl\times\cl$, called {\em maxiconsistent\/} ($\sev$) and {\em weak maxiconsistent
\/} inference ($\wev$).

Our notion of inference is based on maxiconsistency. The idea of using maximal
consistent sets for inference is not new. Maximal consistent sets have been used in
databases (\cite{FUV83}), conditional logic (\cite{RES64}, \cite{VEL76},
\cite{KRA81}, \cite{GIN86}), and belief revision (\cite{AGM85}). However, the notion
of maximal consistency is already present in classical entailment. In order to
compute the inferences of a formula
$\ga$, one can find all maximal consistent sets that do not contain $\neg\ga$, that is
all prime filters containing
$\ga$, and take their intersection. This is the filter that contains all theorems of
$\ga$. Our definition of inference is similar. First, we find all maximal consistent
sets whose elements do not have lower entrenchment than $\neg\ga$.  These sets do not
necessarily contain $\ga$, as opposed, say, to classical logic. Next, we consider
their intersection. If $\ga\lto\gb$ is contained in this intersection then $\ga$
entails maxiconsistently $\gb$, that is $\ga\sev\gb$. It is time to be more formal.

\begin{definition}
Let $U$ be a set of formulas and $\ga\in\cl$. Then the $\ga$-conditionalization
$U^{\ga}$ of
$U$  is the set
$$U^\ga=\{\ga\lto\gb\mid \gb\in U\}.$$
\end{definition}

\begin{lemma}
Let $U,V$ be  deductively closed (under $\vd$). We have the following
\begin{enumerate}
\item $U=V$ iff $U^\ga=V^\ga$.
\item $\Cn (U^\ga,\ga)=\Cn (U,\ga)$.
\end{enumerate}
\end{lemma}

\begin{definition}
Let $\langle \cl, \vd, \prev \rangle$ be an entrenchment frame.
The set of {\em coherent sentences\/}
for a formula $\ga\in\cl$ is the set
$$ \Coh(\ga)=\{\gb\mid \gb\notprev\neg\ga\}.$$
The {\em base\/} of  $\ga$ is the set
$$\cb(\ga)=\{U\mid U=\Cn(U), U\sbs \Coh(\ga) \}.$$
The {\em weak base\/} of  $\ga$ is the set
$$\cb^w(\ga)=\{U\mid U=\Cn(U), U^\ga\sbs \Coh(\ga) \}.$$
The {\em maximal base\/} of $\ga$ is the set
$$\cb_{\max}(\ga)=\{U\mid U \in\cb(\ga)\ \hbox{and if}\ U'=\Cn(U')
\ \hbox{with}\ U\subset U' \ \hbox{then} \ U'\not\in\cb(\ga)\}.$$
The {\em maximal weak base\/} of $\ga$ is the set
$$\cb^w_{\max}(\ga)=\{U\mid U^\ga \in\cb^w(\ga)\ \hbox{and if}\ V=\Cn(V)
\ \hbox{with}\ U^{\ga}\subset V^{\ga} \ \hbox{then} \ V^{\ga}\not\in\cb^w(\ga)\}.$$

The
{\em extension set\/} of $\ga$ is the set
$$e(\ga)=\{\Cn(U,\ga)\mid U\in\cb_{\max}(\ga)\},$$
while the {\em weak extension set\/} of $\ga$ is the set
$$e^w(\ga)=\{U\mid U\in\cb^w_{\max}(\ga)\},$$
The {\em sceptical extension\/} of $\ga$ is the set
$$E(\ga)=\bigcap e(\ga),$$ and
the {\em sceptical weak extension\/} of $\ga$
is the set
$$E^w(\ga)=\bigcap e^w(\ga).$$
Now define
$$\ga\sev\gb\qquad\hbox{iff}\qquad \gb\in E(\ga),$$ and say that {\em $\ga$
maxiconsistently infers
$\gb$ in the entrenchment frame $\langle \cl, \vd, \prev
\rangle$\/}. Also, define
$$\ga\wev\gb\qquad\hbox{iff}\qquad \gb\in E^w(\ga),$$ and say that {\em $\ga$
weak maxiconsistently  infers
$\gb$ in the entrenchment frame $\langle \cl, \vd, \prev
\rangle$\/}.
\end{definition}

Since $\cl$ and $\vd$ remain fixed throughout the following we shall usually drop
$\langle \cl, \vd, \prev \rangle$ and refer to maxiconsistent inference on an entrenchment
consequence relation $\prev$.

Note, that if $\prev$ is $\vd$, that is, if we equate an entrenchment relation with
classical provability then both $\sev$ and $\wev$ collapse to classical $\vd$.

The following lemma deals with inconsistency. In fact, an entrenchment frame is defined in such a
way so that it isolates inconsistency. Also, this lemma ensures that whatever theory
remains after excluding  sets of sentences is consistent. Therefore,
bases and weak bases of a sentence $\ga$ contain only consistent sets with $\ga$.

\begin{lemma}\label{lemma:consistent-theories}
Given an entrenchment frame $\langle
\cl, \vd, \prev \rangle$, the following hold
\begin{enumerate}
\item If $\gb\in \Coh(\ga)$ then $\gb\not\vd\neg\ga$.
\item If $U\sbs \Coh(\ga)$ and $U=\Cn(U)$ then $U,\ga\not\vd\bot$, i.e. $U$ is
consistent with $\ga$.
\item If $U^\ga\sbs \Coh(\ga)$ then $U,\ga\not\vd\bot$, i.e. $U$ is
consistent with $\ga$.\label{weak-consistent-theory}
\item If $U\in \cb^w_{\max}(\ga)$ then $\ga\in U$.\label{weak-contains-it}
\end{enumerate}
\end{lemma}

In the following, we give conditions under which inconsistency is maxiconsistently
derivable.

\begin{lemma}\label{lemma:deriving-inconsistency}
Given an entrenchment frame $\langle
\cl, \vd, \prev \rangle$, then
 $$\ga\sev\bot \quad \hbox{iff}  \quad \Coh(\ga)=\emptyset \quad \hbox{iff}  \quad
\top\prev\neg\ga
\quad \hbox{iff}  \quad   \gb\prev\neg\ga\hbox{, for all $\gb\in\cl$.}$$
\end{lemma}

Note that the above Lemma stiil holds if we replace $\sev$ with $\wev$.

Next we state several properties that  maxiconsistent inference entails in a
entrenchment frame  which will be very useful in the following.

\begin{lemma}\label{lemma:inequalities}
Given an entrenchment frame $\langle
\cl, \vd, \prev \rangle$, the following hold
\begin{enumerate}
\item If $\gb\in \Coh(\ga)$ then $\Cn(\gb)\sbs \Coh(\ga)$.\label{cp-contains-theories}
\item  If $\ga\sev\gb$ then $\ga\lto\neg\gb\prev\neg\ga$.\label{nmr-to-order-one}
\item If $\ga\sev\gb$ then $\neg\gb\prev\neg\ga$.\label{nmr-to-order-two}
\item If $\prev $ is disjunctive then $\ga\prev\gb$ is equivalent to
$\ga\lor\gb\prev\gb$.\label{ld}
\end{enumerate}
\end{lemma}

The corresponding lemma to the above lemma for weak maxiconsistent inference is the
following.

\begin{lemma} \label{lemma:weak-inequalities}
Given an entrenchment frame $\langle
\cl, \vd, \prev \rangle$, the following hold
\begin{enumerate}
\item If $\ga\lto\gb\in \Coh(\ga)$ then $\Cn(\gb)\in
\cb^w(\ga)$.\label{wcp-contains-theories}
\item  If $\ga\wev\gb$ then $\ga\lto\neg\gb\prev\neg\ga$.\label{wnmr-to-order-one}
\item If $\ga\wev\gb$ then $\neg\gb\prev\neg\ga$.\label{wnmr-to-order-two}
\end{enumerate}
\end{lemma}

Bases and weak bases relate to each other through the following lemma.

\begin{lemma}\label{lemma:bases-and-weak-bases}
Given an entrenchment frame $\langle
\cl, \vd, \prev \rangle$, the following hold
\begin{enumerate}
\item If $U\in \cb_{\max}(\ga)$ then $\Cn (U,\ga)\in
\cb^w(\ga)$.\label{base-to-weak-base}
\item If $U\in \cb^w_{\max}(\ga)$ then $\Cn (U^\ga)\sbs
\Coh(\ga)$.\label{weak-base-to-cp}
\end{enumerate}
\end{lemma}

Note that in Part~\ref{base-to-weak-base}, we do {\em not\/} have $\Cn (U,\ga)\in
\cb_{\max}^w(\ga)$. Otherwise, the two notions of maxiconsistent inference would
collapse to each other.

The following lemma shows how different properties of an entrenchment relation
translate to corresponding properties of bases in an entrenchment frame.

\begin{lemma}\label{lem:properties-max-inference} Given an entrenchment frame $\langle
\cl, \vd, \prev \rangle$, the following hold
\begin{enumerate}
\item If  $\ga\vd\gb$ and $\gb\vd\ga$ imply
$\Coh(\ga)=\Coh(\gb)=\Coh(\ga\land\gb)$.\label{logical}
\item If $\prev$ satisfies Bounded  Cut  then $\neg\gb\prev\neg\ga$ implies
$\Coh(\ga)\sbs \Coh(\ga\land\gb)$ (so $\cb(\ga)\sbs\cb(\ga\land\gb)$).\label{cut}
\item If $\prev$ satisfies Bounded Right Monotonicity then $\neg\gb\prev\neg\ga$
implies
$\Coh(\ga\land\gb)\sbs \Coh(\ga)$ (so $\cb(\ga\land\gb)\sbs\cb(\ga)$).\label{cm}
\item If $\prev$ satisfies Bounded Cut and Bounded Right Monotonicity then
$\neg\gb\prev\neg\ga$ implies
$\Coh(\ga\land\gb)=\Coh(\ga)$ (so
$\cb(\ga)=\cb(\ga\land\gb)$).\label{cut-and-cm-with-nmr}
\item If $\prev$ satisfies  Right Monotonicity then $\ga\vd\gb$ implies
$\Coh(\ga)\sbs \Coh(\gb)$ (so $\cb(\ga)\sbs\cb(\gb)$).\label{rm-with-nmr}
\end{enumerate}
\end{lemma}

The corresponding lemma for weak bases is the following.

\begin{lemma}\label{lem:properties-weak-max-inference} Given an entrenchment frame $\langle
\cl, \vd, \prev \rangle$, the following hold
\begin{enumerate}
\item If  $\ga\vd\gb$ and $\gb\vd\ga$ imply
$\cb^w(\ga)=\cb^w(\gb)=\cb^w(\ga\land\gb)$.\label{weak-logical}
\item If $\prev$ satisfies Weak Bounded  Cut  then $\neg\ga\lor\neg\gb\prev\neg\ga$
implies
$\cb^w(\ga)\sbs \cb^w(\ga\land\gb)$.\label{weak-cut}
\item If $\prev$ satisfies Weak Bounded Right Monotonicity then
$\neg\ga\lor\neg\gb\prev\neg\ga$ implies\\
$\cb^w(\ga\land\gb)\sbs \cb^w(\ga)$.\label{weak-cm}
\item If $\prev$ satisfies Weak Bounded Cut and Weak Bounded Right Monotonicity then
$\neg\ga\lor\neg\gb\prev\neg\ga$ implies
$\cb^w(\ga\land\gb)=\cb^w(\ga)$.\label{weak-cut-and-cm-with-entrenchment}
\item If $\prev$ satisfies  Weak Right Monotonicity  then
$\ga\vd\gb$ implies
$\cb^w(\ga)\sbs \cb^w(\gb)$.\label{weak-rm-with-nmr}
\end{enumerate}
\end{lemma}

The following lemmas and theorems are the most important of this section. They
provide us with the converse of Lemma~\ref{lemma:inequalities} and
\ref{lemma:weak-inequalities}. Through these results we are able to reduce the
problem of deciding maxiconsistent inference  in to a problem of deciding an entrenchment
 relation. For that we assume that the entrenchment frame is either disjunctive or
weak disjunctive.

\begin{lemma}\label{lemma:ccf-to-weak}
Let $\prev$ be a weak disjunctive entrenchment relation. Then
\begin{enumerate}
\item $\neg\ga\lor\neg\gb\prev\neg\ga$ iff $\ga\wev\gb$.\label{ccf-to-weak}
\item If $\prev$ satisfies, in addition, Transitivity and Right Conjunction then
$\neg\ga\lor\neg\gb\prev\neg\ga$ implies $\ga\sev\gb$.\label{transitive-ccf-to-weak}
\end{enumerate}
\end{lemma}

Given the above lemma, we can state the connection between maxiconsistent and weak
maxiconsistent inference on a weak disjunctive entrenchment frame.

\begin{theorem}\label{theorem:wd-sev-and-wev}
Let $\prev$ be a weak disjunctive entrenchment relation. Then
\begin{enumerate}
\item $\ga\sev\gb$ implies $\ga\wev\gb$. \label{sev-to-wev}
\item If $\prev$ satisfies, in addition, Transitivity and Right Conjunction then
$\ga\wev\gb$ implies $\ga\sev\gb$.\label{transitive-wev-to-sev}
\end{enumerate}
\end{theorem}

The following theorem shows that, for disjunctive entrenchment relations,
 maxiconsistent inference is decided by a kind of contraposition. We have that
$\ga$ maxiconsistently infers $\gb$ if ``$\neg\gb$ is less entrenched than
$\neg\ga$''.

\begin{theorem}\label{thm:ccf-to-strong}
Let $\prev$ be a disjunctive entrenchment relation. Then
 $$\neg\gb\prev\neg\ga \quad\hbox{if and only if}\quad \ga\sev\gb.$$
\end{theorem}

The following corollary says that maxiconsistent and weak maxiconsistent inference
coincide on disjunctive entrenchment frames.

\begin{corollary}\label{coroll:in-disjunctive-weak-equals-strong}
Let $\prev$ be a disjunctive entrenchment relation. Then
 $$\ga\sev\gb\quad\hbox{if and only if}\quad \ga\wev\gb.$$
\end{corollary}

\section{Nonmonotonic Consequence Relations and their Representations}
\label{sec:nmir}

A recent breakthrough in nonmonotonic logic is the beginning of study of nonmonotonic
consequence through postulates for abstract nonmonotonic consequence relations, using
Gentzen-like context-sensitive sequents (\cite{GAB85},
\cite{MAK89}, \cite{KLM90}). The outcome of this research turns out to be valuable in
at least two ways

\begin{itemize}

\item it provides a sufficiently general {\em axiomatic\/} framework for comparing and
classifying nonmonotonic formalisms, and

\item it gave rise to new, simpler, and better behaved systems for nonmonotonic
reasoning, such as cumulative (\cite{GAB85}), preferential (\cite{KLM90}), and
rational (\cite{LM92}) inference relations.

\end{itemize}

In this paper, we shall present a variety of representations results for nonmonotonic
consequence relations through maxiconsistent inference on entrenchment frames.

Before presenting  the  results of this section (and main results of this paper), we
shall define a variety of classes of nonmonotonic consequence relations.   The rules
mentioned in the following are presented in Table~\ref{table:rules}. For a motivation
of these rules see
\cite{KLM90} and
\cite{MAK94}. (The latter serves as an excellent introduction to nonmonotonic
consequence relations.)

\begin{table}[h]
\begin{center}
\begin{minipage}{4.5in}
{\small
\begin{tabular}{cl}

$\displaystyle \frac{\ga \vd \gb}{\ga\ev\gb}$ & {\footnotesize (Supraclassicality)}
\vspace{.1in} \\

$\displaystyle \frac{\ga\vd\gb \quad \gb\vd\gc \quad\ga\ev\gc}{\gb\ev\gc}$ &
{\footnotesize (Left Logical Equivalence)} \vspace{.1in}\\

$\displaystyle \frac{\ga\ev\gb \qquad \gb\vd\gc}{\ga\ev\gc}$ &
{\footnotesize(Right Weakening)} \vspace{.1in}\\

$\displaystyle \frac{\ga\ev\gb \qquad \ga\ev\gc}{\ga\ev\gb\land\gc}$ &
{\footnotesize(And)} \vspace{.1in}\\

$\displaystyle \frac{\ga\ev\gb \qquad
\ga\land\gb\ev\gc}{\ga\ev\gc}$ & {\footnotesize(Cut)} \vspace{.1in}\\

$\displaystyle \frac{\ga\ev\gb \qquad
\ga\ev\gc}{\ga\land\gb\ev\gc}$ & {\footnotesize(Cautious
Monotonicity)}\vspace{.1in}\\

$\displaystyle \frac{\ga_0\ev\ga_1\quad\cdots\quad \ga_{n-1}\ev\ga_n\quad
\ga_n\ev\ga_0}{\ga_0\ev\ga_n}$ & {\footnotesize(Loop)}\vspace{.1in}\\

$\displaystyle \frac{\ga\ev\gc \qquad
\gb\ev\gc}{\ga\lor\gb\ev\gc}$ & {\footnotesize(Or)} \vspace{.1in}\\

$\displaystyle \frac{\ga\lor\ev\gb \qquad
\gb\lor\gc\ev\gb}{\ga\lor\gc\ev\ga}$ & {\footnotesize(Weak Transitivity)}
\vspace{.1in}\\

$\displaystyle \frac{\ga\notev\neg\gb \qquad
\ga\ev\gc}{\ga\land\gb\ev\gc}$ & {\footnotesize(Rational Monotonicity)}
\end{tabular} }
\end{minipage}

\end{center}

\caption{\label{table:rules} Rules for Nonmonotonic Inference}
\end{table}

\begin{definition} Following (\cite{KLM90}, \cite{LM92}, \cite{GM94}), we shall say
that a relation $\ev$ on $\cl$ is a {\em nonmonotonic consequence relation (based on
$\vd$)\/} if it satisfies Supraclassicality, Left Logical Equivalence, Right
Weakening, and And.  We  call a nonmonotonic consequence relation $\ev$ {\em
cumulative\/} if it satisfies, in addition, Cut and Cautious Monotonicity, {\em
strongly cumulative\/} if it is cumulative and satisfies, in addition, Loop,  {\em
preferential\/} if it is cumulative and satisfies, in addition,  Or, and {\em
rational\/} if it is preferential and satisfies, in addition,  Rational Monotonicity.
\end{definition}

The most controversial of these rules is Rational Monotony, which, moreover, is
non-Horn. For a plausible counterexample, see~\cite{STA94}.

The class of nonmonotonic consequence relations is too general and therefore very weak.
The class of default inference relations contains sceptical inference of Reiter's
default systems~\cite{REI80}. Poole systems with constraints (\cite{POO88}) and
cumulative  default systems such as the one appeared in~\cite{BREW91}  belong to the
class of cumulative inference relations. Strong cumulativity has no concrete
formalism, as far as we know. Inference defined on Poole systems without constraints
as well as entailment on classical preferential models  belong to the class of
preferential inference relations. Finally, ranked operators (\cite{KG95c}), as well
as, the AGM belief revision operator  belong to the class of rational
inference relations.

\subsection{ Maxiconsistent Inference}

The first theorem of this section shows that  maxiconsistent inference in an
arbitrary entrenchment frame is a nonmonotonic consequence relation. All subsequent results
assume that the entrenchment frame is either disjunctive or weak disjunctive.

\begin{theorem}\label{thm:soundness}
Let $\langle \cl, \vd, \prev \rangle$ be a
entrenchment frame. Then its maxiconsistent inference
$\sev$ is a nonmonotonic consequence relation. Moreover,
\begin{enumerate}
\item If $\prev$ satisfies Bounded Cut and Bounded Right Monotonicity then $\sev$ is a
cumulative inference relation.\label{sound-cumulative}
\item If $\prev$ satisfies  Bounded Cut and Right Monotonicity  then $\sev$ is a
strong cumulative inference relation.\label{sound-strong-cumulative}
\item If $\prev$ satisfies Transitivity and Right Conjunction then $\sev$ is a
preferential inference relation.\label{sound-preferential}
\end{enumerate}
\end{theorem}

From now on, we will assume a disjunctive entrenchment relation. The
maxiconsistent inference of a disjunctive entrenchment relation will give a
canonical representation of nonmonotonic consequence relation. The following
definitions provide, for each  nonmonotonic consequence relation,  an entrenchment
 relation with the same  maxiconsistent inference, and conversely.

\begin{definition} Given an entrenchment relation $\prev$ and a nonmonotonic  inference
relation $\ev$, then define a consequence relation $\ev'$ and a relation $\prev'$ as
follows
\\
\begin{tabular}{lrcl}
($N$) & $\ga\ev'\gb$ & iff &  $\neg\gb\prev\neg\ga$\\
($P$) & $\ga\prev'\gb$ & iff & $\neg\gb\ev\neg\ga$.\\
\end{tabular}
\\
We shall also denote $\ev'$ and $\prev'$ with $N(\prev)$ or
 and $P(\ev)$, respectively.
\end{definition}

Given the above definition one can prove the following lemma
\begin{lemma}\label{lemma:iso-entrenchment}
Let $\prev$ and $\ev$ be an entrenchment  and  a nonmonotonic
consequence relation, respectively. Then
\begin{enumerate}
\item $P(N(\prev))=\prev$, and
\item $N(P(\ev))=\ev$.\label{ev-to-ev}
\end{enumerate}
\end{lemma}

\begin{corollary}\label{coroll:definition-equals-strong}
Let $\prev$ be a disjunctive entrenchment relation. Then
$$N(\prev)=\sev,$$
where $\sev$ is the  maxiconsistent inference of $\prev$.
\end{corollary}

We have the following
\begin{theorem}\label{theorem:disjunctive-soundness}
Let $\langle \cl, \vd, \prev \rangle$ be a
disjunctive entrenchment frame. Then
the inference relation $\ev$ defined by $N$ is a nonmonotonic consequence relation such
that, for all
$\ga$, $\gb$ in $\cl$,
$$\ga\ev\gb \qquad\hbox{iff}\qquad \ga\sev\gb.$$
Moreover,
if $\prev$ satisfies Bounded Cut, Bounded Right Monotonicity, Acyclicity and
Conjunction  then
$\sev$ satisfies Cut, Cautious Monotonicity, Loop and Or,
respectively.
\end{theorem}

Going from nonmonotonic consequence relations to disjunctive entrenchment
relations, we have the following theorem.

\begin{theorem}
\label{thm:completeness-cut-cm}
Let $\ev$ be a nonmonotonic inference
relation, then the relation $\prev$ defined by
$(P)$ is  a disjunctive entrenchment relation
such that, for all
$\ga$, $\gb$ in $\cl$,
$$\ga\ev\gb \qquad\hbox{iff}\qquad \ga\sev\gb.$$
Moreover,
if $\ev$ satisfies  Cut, Cautious Monotonicity, Loop, and Or then $\prev$ satisfies
Bounded Cut, Bounded Right Monotonicity, Acyclicity, and Conjunction, respectively.
\end{theorem}

\subsection{Weak Maxiconsistent Inference}

In this section, we will study weak maxiconsistent inference on weak disjunctive entrenchment
frames. This will allow us to find  better behaved entrenchment
relations equivalent with a given nonmonotonic consequence relation. First, a theorem
analogous to Theorem~\ref{thm:soundness}
 which shows that weak maxiconsistent inference is nonmonotonic.

\begin{theorem}\label{thm:weak-soundness}
Let $\langle \cl, \vd, \prev \rangle$ be a
entrenchment frame. Then its weak maxiconsistent inference
$\wev$ is a nonmonotonic consequence relation.
\end{theorem}

As in the previous section, we will define maps between the classes of nonmonotonic
consequence relations and weak disjunctive entrenchment relations, and
conversely.
\begin{definition} Given an entrenchment relation $\prev$ and a nonmonotonic  inference
relation $\ev$, then define a consequence relation $\ev'$ and a relation $\prev'$ as
follows
\\
\begin{tabular}{rrcl}
($N_{\lto}$) & $\ga\ev'\gb$ & iff & $\neg\ga\lor\neg\gb\prev\neg\ga$\\
($P_{\lto}$) & $\ga\prev'\gb$ & iff & $\neg\ga\lor\neg\gb\ev\neg\ga$.\\
($P_{tr}$)   & $\ga\prev''\gb$ & iff & there exist $\gd_1,\ldots,\gd_n\in\cl$  such
                                       that \\
              &               &       &  $\neg\gb\ev\gd_1, \gd_1\ev\gd_2, \ldots,
                                       \gd_n\ev\neg\ga$.
\end{tabular}
\\
We shall also denote $\ev'$, $\prev'$ and $\prev''$ with
$N_{\lto}(\prev)$,
$P_{\lto}(\ev)$ and $P_{tr}(\ev)$, respectively.
\end{definition}

Given the above definition one can prove the following lemma

\begin{lemma}\label{lemma:weak-iso-entrenchment}
Let $\prev$ be an entrenchment relation and $\ev$ a nonmonotonic consequence relation.
Then
\begin{enumerate}
\item  if $\prev $ satisfies Right Monotonicity and Right Conjunction then\\
$P_{\lto}(N_{\lto}(\prev))=\prev$,\label{lto-prev}
\item $N_{\lto}(P_{\lto}(\ev))=\ev$,\label{lto-ev}
\item if $\prev$ is transitive then $P_{tr}(N_\lto(\prev))=\prev$, and
\label{tr-prev}
\item if $\ev$ satisfies Loop then $N_\lto(P_{tr}(\ev))=\ev$.\label{tr-ev}
\end{enumerate}
\end{lemma}

\begin{corollary}\label{coroll:definition-equals-weak}
Let $\prev$ be a weak disjunctive entrenchment relation. Then
$$N_\lto(\prev)=\wev,$$
where $\wev$ is the weak maxiconsistent inference of $\prev$.
\end{corollary}

We now have the following
\begin{theorem}\label{theorem:weak-completeness}
Let $\langle \cl, \vd, \prev \rangle$ be a weak
disjunctive entrenchment frame. Then
the inference relation $\ev$ defined by $N_\lto$ is a nonmonotonic consequence relation
such that, for all
$\ga$, $\gb$ in $\cl$,
$$\ga\ev\gb \qquad\hbox{iff}\qquad \ga\wev\gb.$$
Moreover,
if $\prev$ satisfies Weak Bounded Cut, Weak Bounded Right Monotonicity, Weak
Acyclicity and Right Conjunction  then
$\wev$ satisfies Cut, Cautious Monotonicity, Loop and Or,
respectively.
\end{theorem}

We do not have a similar theorem to Theorem~\ref{thm:completeness-cut-cm} because an
arbitrary nonmonotonic consequence relation does not define an entrenchment
relation through ($P_\lto$). However, it does so if we assume that it is {\em
preferential\/}.

\begin{theorem}
\label{thm:completeness-preferential}
Let $\ev$ be a preferential inference
relation, then the relation $\prev$ defined by
$(P_\lto)$ is  a weak disjunctive and transitive entrenchment relation
satisfying Conjunction
such that, for all
$\ga$, $\gb$ in $\cl$,
$$\ga\ev\gb \qquad\hbox{iff}\qquad \ga\sev\gb\qquad\hbox{iff}\qquad\ga\wev\gb.$$
\end{theorem}

We can characterize  strong cumulative inference relations through weak
maxiconsistent inference, if we employ ($P_{tr}$).

\begin{theorem}
\label{thm:completeness-strong-cumulative}
Let $\ev$ be a nonmonotonic consequence relation satisfying Loop, then    the
 relation
$\prev$ defined by ($P_{tr}$) is a  weak disjunctive transitive entrenchment
relation such that, for all
$\ga$, $\gb$ in $\cl$,
$$\ga\ev\gb \qquad\hbox{iff}\qquad \ga\wev\gb.$$
\end{theorem}

\section{Conclusion}\label{section:summary}

In this section, we will give a summary of the correspondence between classes of
entrenchment and nonmonotonic consequence relations.

Let $\ca$ be a class of nonmonotonic
consequence relations and $\cb$ a class of
entrenchment relations. Let ${\sf C},{\sf C}^w$ be  maps from $\cb $ to $\ca$
with
${\sf C}(\prev)=\sev$ and ${\sf C}^w(\prev)=\wev$, respectively, where $\sev$ an
$\wev$ are the maxiconsistent and weak maxiconsistent inference on $\prev$.

We
will say that a class
$\ca$ of nonmonotonic consequence relations is {\em dual\/} to a class $\cb$ of
entrenchment relations and denote it with $\ca\equiv\cb$ if there exists a map
$\sn$ such that
$\sn:\cb\lto\ca$,
$\sfc\circ \sn=\hbox{\sf Id}_\ca$, and
$\sn\circ \sfc=\hbox{\sf Id}_\cb$, where $\hbox{\sf Id}$ is the identity map.
Similarly, $\ca$ and
$\cb$ will be {\em weakly dual\/} and we denote it with
$\ca\stackrel{w}{\equiv}\cb$ if there exists a map
$\sn$ such that
$\sn:\cb\lto\ca$,
$\sfc^w\circ \sn=\hbox{\sf Id}_\ca$, and
$\sn\circ \sfc^w=\hbox{\sf Id}_\cb$.

We
will say that a class
$\ca$ of nonmonotonic consequence relations is {\em a retract\/} of a class $\cb$ of
entrenchment relations and denote it with $\ca\mid\!\equiv\cb$ if there exists
a map
$\sn$ such that
$\sn:\cb\lto\ca$ and
$\sfc\circ \sn=\hbox{\sf Id}_\ca$. Similarly,
$\ca$  is {\em a weak retract\/} of  $\cb$  and we denote it with
$\ca\mid\!\stackrel{w}{\equiv}\cb$ if there exists a map
$\sn$ such that
$\sn:\cb\lto\ca$ and
$\sfc^w\circ \sn=\hbox{\sf Id}_\ca$.

A list of all classes of nonmonotonic and entrenchment relations mentioned in
the following appear on Table~\ref{table:classes}

\begin{table}[h]
\begin{center}
\begin{minipage}{4.5in}
{\small
\begin{tabular}{lcl}
$\bf{NM}$ & = & all nonmonotonic consequence relations (nmcr)\\
$\bf{D}$ & = & nmcr satisfying Cut\\
$\bf{CM}$ & = & nmcr satisfying Cautious Monotonicity\\
$\bf{C}$ & = &  cumulative nmcr\\
$\bf{SC}$ & = &  strong cumulative nmcr\\
$\bf{P}$ & = & preferential nmcr \\
$\bf{E}$ & = & all entrenchment relations (er)\\
$\bf{BC}$ & = & er satisfying Bounded Cut\\
$\bf{BR}$ & = & er satisfying Bounded Right Monotonicity\\
$\bf{BCR}$ & = & er satisfying Bounded Cut and Bounded Right Monotonicity\\
$\bf{BA}$ & = & er satisfying Bounded Cut, Bounded Right Monotonicity,\\
          &   &  and Acyclicity\\
$\bf{T}$ & = & er satisfying Transitivity\\
$\bf{TC}$ & = & er satisfying Transitivity and Right Conjunction\\
$\bf{d}\hbox{-}\cb$ & = & er satisfying the properties of $\cb$ and Left Disjunction\\
$\bf{wd}\hbox{-}\cb$ & = & er satisfying the properties of $\cb$ and Weak Left
Disjunction
\end{tabular}
}
\end{minipage}
\end{center}
\caption{\label{table:classes} Classes of nonmonotonic and entrenchment
relations}
\end{table}

The classes of nonmonotonic consequence relations relate to each other through the
following scheme (right to left direction denotes inclusion).
\vspace{.1in}

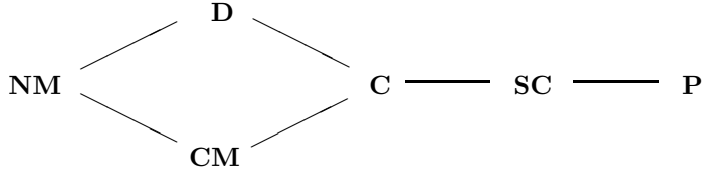
\setlength{\unitlength}{0.00083300in}%
\begin{picture}(4200,1035)(1426,-1438)
\thinlines
\put(1876,-811){\line( 2, 1){600}}
\put(1876,-961){\line( 2,-1){600}}
\put(4951,-886){\line( 1, 0){525}}
\put(3901,-886){\line( 1, 0){525}}
\put(2941,-1291){\line( 2, 1){600}}
\put(2950,-493){\line( 2,-1){600}}
\put(2551,-1411){$\bf CM$}
\put(5626,-961){$\bf P$}
\put(4576,-961){$\bf SC$}
\put(3676,-961){$\bf C$}
\put(1426,-961){$\bf NM$}
\put(2686,-511){$\bf D$}
\end{picture}

Similarly, the classes of entrenchment relations relate to each other
as follows.
\vspace{.1in}

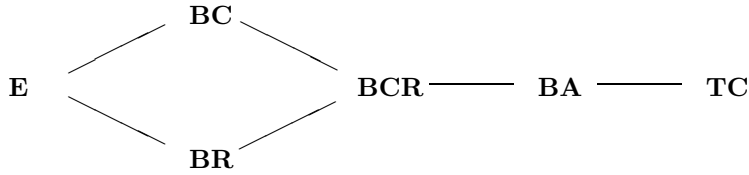
\setlength{\unitlength}{0.00083300in}%
\begin{picture}(4350,1035)(1501,-1438)
\thinlines
\put(1876,-811){\line( 2, 1){600}}
\put(1876,-961){\line( 2,-1){600}}
\put(2941,-1291){\line( 2, 1){600}}
\put(2950,-493){\line( 2,-1){600}}
\put(4126,-886){\line( 1, 0){525}}
\put(5176,-886){\line( 1, 0){525}}
\put(3676,-961){$\bf BCR$}
\put(2626,-511){$\bf BC$}
\put(2626,-1411){$\bf BR$}
\put(4801,-961){$\bf BA$}
\put(5851,-961){$\bf TC$}
\put(1501,-961){$\bf E$}
\end{picture}

Moreover, If $\cb$ is any entrenchment relation class then
$\bf{d}\hbox{-}\cb$ and $\bf{wd}\hbox{-}\cb$ are $\cb$ augmented with Left
Disjunction and Weak Left Disjunction, respectively. Clearly,
$\bf{d}\hbox{-}\cb \sbs \bf{wd}\hbox{-}\cb \sbs \cb$.

We now have the following corollary
\begin{corollary}\label{coroll:classes}
The following hold
\begin{enumerate}

\item ${\bf NM}\equiv \bf{d}\hbox{-}\bf E$,
${\bf NM}\stackrel{w}{\equiv} \bf{d}\hbox{-}\bf E$,
${\bf NM}\mid\!\equiv\bf{E}$,
${\bf NM}\mid\!\stackrel{w}{\equiv}\bf{E}$, and
${\bf NM}\mid\!\equiv\bf{wd}\hbox{-}\bf{E}$.\label{NM}

\item ${\bf D}\equiv \bf{d}\hbox{-}\bf{BC}$ and
${\bf D}\stackrel{w}{\equiv}\bf{d}\hbox{-}\bf{BC}$.\label{D}

\item ${\bf CM}\equiv \bf{d}\hbox{-}\bf{BR}$ and
${\bf CM}\stackrel{w}{\equiv}\bf{d}\hbox{-}\bf{BR}$.\label{CM}

\item ${\bf C}\equiv \bf{d}\hbox{-}\bf{BCR}$,
${\bf C}\stackrel{w}{\equiv}\bf{d}\hbox{-}\bf{BCR}$,
${\bf C}\mid\!\equiv\bf{BCR}$ and
${\bf C}\mid\!\stackrel{w}{\equiv}\bf{BCR}$.\label{C}

\item ${\bf SC}\equiv \bf{d}\hbox{-}\bf{BA}$,
${\bf SC}\stackrel{w}{\equiv}\bf{d}\hbox{-}\bf{BA}$ and
${\bf SC}\stackrel{w}{\equiv}\bf{wd}\hbox{-}\bf{T}$.\label{SC}

\item ${\bf P}\mid\!\equiv \bf{d}\hbox{-}\bf{TC}$,
${\bf P}\equiv\bf{wd}\hbox{-}\bf{TC}$ and
${\bf P}\stackrel{w}{\equiv}\bf{wd}\hbox{-}\bf{TC}$.\label{P}

\end{enumerate}
\end{corollary}

\end{document}